\documentclass[iop,revtex4]{emulateapj}

\usepackage{apjfonts}
\usepackage{times}

\usepackage{amsmath}
\usepackage{graphicx}
\usepackage{natbib}
\usepackage{epstopdf} %% for pdflatex
\usepackage{subfigure}
\usepackage{color}
\usepackage{lscape}

 \font\sevenrm=cmr7 scaled 1000
\newcommand{\mbh}{$M_{\rm BH}$}     
     
\newcommand{\ergs}{erg~s$^{\rm -1}$}
       
\newcommand{\CIV}{C~{\sevenrm IV}}
\newcommand{\NV}{N~{\sevenrm V}}
\newcommand{\NIV}{N~{\sevenrm IV}]}
\newcommand{\OIII}{[O~{\sevenrm III}]}
\newcommand{\OIV}{O~{\sevenrm IV}]}
\newcommand{\HeII}{He~{\sevenrm II}}

\newcommand{\SiIV}{Si~{\sevenrm IV}}

\newcommand{\SII}{[S~{\sevenrm II}]}
\newcommand{\OI}{[O~{\sevenrm I}]}
\newcommand{\OII}{[O~{\sevenrm II}]}
\newcommand{\NII}{[N~{\sevenrm II}]}
\newcommand{\Hb}{H$\beta$}
\newcommand{\Ha}{H$\alpha$}
\newcommand{\MgII}{Mg~{\sevenrm II}}
\begin{document}
\title{
Outflow and metallicity in the broad-line region of low-redshift active galactic nuclei}
\author{Jaejin Shin$^{1}$}
\author{Tohru nagao$^{2}$}
\author{Jong-Hak Woo$^{1}$\altaffilmark{,3}}

\affil{
$^1$Astronomy Program, Department of Physics and Astronomy, 
Seoul National University, Seoul, 151-742, Republic of Korea\\
$^2$Research Center for Space and Cosmic Evolution, Ehime University, 
Bunkyo-cho 2-5, Matsuyama, Ehime 790-8577, Japan
}
\altaffiltext{3}{Author to whom any correspondence should be addressed}

\begin{abstract}
Outflows in active galactic nuclei (AGNs) are crucial to understand in
investigating the co-evolution of supermassive black holes (SMBHs) and their host
galaxies since outflows may play an important role as an AGN feedback mechanism.
Based on the archival UV spectra obtained with HST and IUE,
we investigate outflows in the broad-line region (BLR) in low-redshift AGNs ($z < 0.4$)
through the detailed analysis of the velocity profile of the \CIV\ emission line. We find a
dependence of the outflow strength on the Eddington ratio and the BLR
metallicity in our low-redshift AGN sample, which is consistent with the earlier
results obtained for high-redshift quasars. These results suggest that
the BLR outflows, gas accretion onto SMBH, and past star-formation activity
in the host galaxies are physically related in low-redshift AGNs as in 
powerful high-redshift quasars.
\end{abstract}


%%Section 1
\section{INTRODUCTION} \label{section:intro}

It is widely believed that a supermassive black hole (SMBH) resides at
the central part of most massive galaxies, and the mass of SMBHs 
($M_{\rm BH}$) reaches up to $\sim$$10^{10} M_\odot$ (e.g., 
\citealt{Vestergaard2008, Schulze2010}; see also \citealt{Wu2015}).
Interestingly, scaling relations have been observed 
between the SMBH mass and physical properties of their host galaxies,
regardless of AGN activity
%, i.e., the bulge mass, stellar velocity dispersion, and infrared luminosity 
\citep[e.g.,][]{Magorrian1998, Ferrarese2000, Gebhardt2000, Marconi2003, 
Haring2004, Kormendy2013, Woo2013, Woo2015}, suggesting
that SMBHs and galaxies evolved with a close interaction 
(e.g., \citealt{Merloni2004}). However, the physics 
behind the co-evolution is unclear, preventing us from full 
understandings of the cosmological evolution of galaxies and SMBHs.

The gas accretion onto SMBHs is a fundamental process to explain the origin 
of the huge luminosity of active galactic nuclei (AGNs). Since AGNs are 
often associated with the star formaiton in their host galaxies (e.g., 
\citealt{Heckman1997,CidFernandes2001,Netzer2009,Woo2012,Matsuoka2015}), AGNs 
are a key population to explore the SMBH-galaxy co-evolution. More recent 
works suggest that AGN activity is important to quench 
star formation in their host galaxies (the negative AGN feedback; 
see, e.g., \citealt{Kauffmann2000, Granato2004, Dimatteo2005, Croton2006,
Hopkins2006, Ciotti2010, Scannapieco2012}; cf. for positive AGN feedback, 
see, e.g., \citealt{Silk2005, Gaibler2012, Zubovas2013, Ishibashi2013}) then 
the quenching of the star-forming activity is expected to terminate the AGN
activity. Therefore the AGN feedback is now regarded as a crucial process 
for the SMBH-galaxy co-evolution.

Outflows are often considered as one of the AGN feedback mechanisms.
The velocity profile and shift from the systemic velocity of various 
absorption and emission lines observed in UV and optical spectra of AGNs 
suggest the presence of the strong outflow of ionized gas, that exist in both small 
spatial scales corresponding to the broad-line region (BLR, that is located at 
the sub-pc scale from the nucleus) and large spatial scales corresponding to 
the narrow-line region (NLR, that is located at $\sim$$10^{1-4}$ pc from the 
nucleus) \citep{Weymann1991, Crenshaw2003, Ganguly2007,Mueller2011, 
Harrison2014, Bae2014, Husemann2016, Woo2016, Karouzos2016, Woo2017}. 
Not only the ionized gas, powerful outflows of molecular 
gas at the scale of their host galaxies are also seen in sub-millimeter and millimeter 
spectra \citep[e.g.,][]{Maiolino2012, Garcia2014}. The inferred kinetic 
power of AGN outflows is high enough ($\sim 10^{43-45}$ erg s$^{-1}$; e.g., 
\citealt{Tombesi2012}) that a large amount of ISM in AGN host 
galaxies can be blown away, resulting in the termination of the star-formation 
activity. However, the detailed properties and physical origin of the AGN 
outflow are not understood well, and thus further observational studies on the 
AGN outflow is crucial to reveal the nature of the AGN feedback.

Though the AGN outflow is seen in various spatial scales as described above,
the BLR has been particularly investigated to examine the nature of the AGN 
outflow. Especially high-ionization emission lines from BLRs such as 
\CIV$\lambda$1549 are interesting for assessing the BLR outflow. This is 
because those high-ionization lines are emitted from the closer region to the 
AGN central engine than low-ionization lines, such as \MgII, as suggested by 
reverberation mapping observations \citep[e.g.,][]{Clavel1991,Sulentic2000,
Wang2012} and also by photoionization models \citep[e.g.,][]{Baldwin1995,
Korista2000}. In fact, the \CIV\ emission line sometimes show significant 
blueshifts \citep{Gaskell1982, Wilkes1984, Marziani1996, Sulentic2000, 
Vandenberk2001, Richards2002, Baskin2005, Sulentic2007} or asymmetric 
velocity profiles \citep{Sulentic2000, Baskin2005, Sulentic2007} than 
low-ionization BLR lines. 
% Of interest, the blueshift and asymmetric features in the \CIV\ velocity profile 
% are preferably seen for radio-quiet and radio-loud QSOs respectively, even 
% though its physical origin is unknown (e.g., Sulentic et al. 2000). 
It has been also reported that the outflow properties in BLRs are also related to 
the Eddington ratio \citep[e.g.,][]{Wang2011}. In terms of the SMBH-galaxy 
co-evolution, the relation between the AGN outflow properties and the metallicity 
is specifically important because the metallicity is strongly linked to the past 
star-formation history of the host galaxy. \cite{Wang2012} reported the close 
relation between the BLR metallicity ($Z_{\rm BLR}$) and AGN outflow through 
the analysis of the \CIV\ emission-line profile in a SDSS high-$z$ ($1.7 < z < 
4.0$) QSO sample, and concluded that the past star formation in galaxy could 
affect both the accretion activity and the AGN outflow in the BLR scale. 

There is a limitation in the observational studies of the AGN outflows; the BLR 
outflow has been explored often for high-redshift QSOs 
\citep[e.g.,][]{Crenshaw2003, Wang2011, Wang2012}. This is because 
high-ionization BLR lines such as \CIV\ are in the rest-frame UV and thus they 
are not observable from ground-based telescopes. The lack of systematic 
studies of the BLR outflow at low-redshift Universe prevents us from examining 
the redshift evolution of the outflow, that is crucial to understand the role of AGN 
feedback within the context of the SMBH-galaxy co-evolution in the cosmological 
timescale. Another motivation for the BLR outflow study in low-redshift AGNs is 
from the expectation that the basic properties and physics of the AGN outflow 
could be different between high redshift and low redshift, since the typical gas 
fraction of host galaxies is expected to be completely different between 
high-redshift and low-redshift AGNs \citep{Daddi2010,Tacconi2010,Geach2011,Bauermeister2013,Tacconi2013,Popping2015}.

\cite{Shin2013} presented rest-frame UV spectra of 70 low-redshift
($z < 0.5$) PG QSOs for investigating the relation between $Z_{\rm BLR}$
and AGN properties such as the SMBH mass, AGN luminosity, and
Eddington ratio. This sample is extremely useful also for studying the nature of
the BLR outflow in low-redshift AGNs, especially for examining possible 
relations between the outflow activity and the metallicity in the BLR scale.
Therefore, in this paper, we investigate the BLR outflow for the sample of PG
QSOs given in \cite{Shin2013}. The structure of this paper is as follows.
We describe the sample selection and the data in \S 2, and then explain the 
data analysis especially on the determination of some outflow indicators in \S 3.
The main results are presented in \S 4, followed by the discussion in \S 5. 
Finally, the summary and conclusions are given in \S 6. We adopt a cosmology 
of $H_{\rm 0}= 70$ km  s$^{-1}$ Mpc$^{-1}$,  
$\Omega_{\Lambda}=0.7$ and $\Omega_{\rm m}=0.3$.\\

\begin{deluxetable}{lclr} 
\tablewidth{0pt}
\tablecolumns{4}
\tabletypesize{\scriptsize}
\tablecaption{The archival UV data used in this study}

\tablehead{
\colhead{Target ID} &
\colhead{Redshift} &
\colhead{Observation Date} &
\colhead{Telescope/Instrument}
\\
\colhead{(1)} &
\colhead{(2)} &
\colhead{(3)} &
\colhead{(4)} 
}

\startdata
Mrk~106		& $0.123$  &2011 May 12, 13		&HST/COS		 \\
Mrk~110		& $0.035$  &1988 Feb 28			&IUE/SWP		 \\
Mrk~290		& $0.030$  &2009 Oct 28			&HST/COS		 \\
Mrk~493		& $0.031$  &1996 Sep 04			&HST/FOS		 \\
Mrk~506		& $0.043$  &1979 Jul 03			&IUE/SWP		 \\
Mrk~1392		& $0.036$  &2004 Jun 07			&HST/STIS		\\
PG~0157+001	& $0.164$  &1985 Aug 09			&IUE/SWP		 \\
PG~0921+525	& $0.035$  &1988 Feb 28,29		&IUE/SWP		 \\
PG~0923+129	& $0.029$  &1985 May 01			&IUE/SWP		 \\
PG~0947+396	& $0.206$  &1996 May 06			&HST/FOS		 \\
PG~1022+519	& $0.045$  &1983 May 31; Jun 01	&IUE/SWP		 \\
PG~1048+342	& $0.167$  &1993 Nov 13			&IUE/SWP		 \\
PG~1049-005	& $0.357$  &1992 Apr 01,			&HST/FOS		 \\
PG~1115+407	& $0.154$  &1996 May 19			&HST/FOS		 \\
PG~1121+422	& $0.234$  &1995 Jan 08			&IUE/SWP		 \\
PG~1151+117	& $0.176$  &1987 Jan 29, 30		&IUE/SWP		 \\
PG~1202+281	& $0.165$  &1996 Jul 21			&HST/FOS		 \\
PG~1229+204	& $0.063$  &1982 May 01, 02		&IUE/SWP		 \\
PG~1244+026	& $0.048$  &1983 Feb 08			&IUE/SWP		 \\
PG~1307+085	& $0.155$  &1980 May 04			&IUE/SWP		\\
PG~1341+258	& $0.087$  &1995 Mar 22			&IUE/SWP		 \\
PG~1404+226	& $0.098$  &1996 Feb 23			&HST/FOS		 \\
PG~1415+451	& $0.114$  &1997 Jan 02 			&HST/FOS		 \\
PG~1425+267	& $0.366$  &1996 Jun29			&HST/FOS		 \\
PG~1427+480	& $0.221$  &1997 Jan 07			&HST/FOS		 \\
PG~1444+407	& $0.267$  &1996 May 23			&HST/FOS		 \\
PG~1448+273	& $0.065$  &2011 Jun 18	 	 	&HST/COS		\\
PG~1512+370	& $0.371$  &1992 Jan 26			&HST/FOS		 \\
PG~1519+226	& $0.137$  &1995 Jun 11			&IUE/SWP		 \\
PG~1534+580	& $0.030$  &2009 Oct 28			&HST/COS		 \\
PG~1545+210	& $0.266$  &1992 Apr 08, 10		&HST/FOS		 \\
PG~1552+085	& $0.119$  &1986 Apr 28			&IUE/SWP		 \\
PG~1612+261	& $0.131$  &1980 Sep 10			&IUE/SWP		 \\
PG~2233+134	& $0.325$  &2003 May 13			&HST/STIS			 
\enddata
\label{table:prop}

\tablecomments{
    Col. (2): Redshift adopted from the NED.}
\end{deluxetable}

%section 2
\section{Sample Selection and the Data} \label{section:sample}

As mentioned in \S 1, high-ionization emission lines from the BLR seen in the
rest-frame UV spectrum are generally used to study the AGN outflow in the sub-pc scales. 
We focus on the \CIV\ line in this work, since other high-ionization 
BLR lines are heavily blended with other lines (e.g., \NV$\lambda$1240, 
\OIV$\lambda$1402) or too weak to investigate in detail (e.g., 
\NIV$\lambda$1486, \HeII$\lambda$1640). In addition to \CIV, optical 
emission lines from the NLR are necessary in the outflow analysis to define 
the reference redshift (see \S 3.1 for more details). Therefore we need a 
sample of AGNs whose UV and optical spectra with a high signal-to-noise ratio 
are available.

In \cite{Shin2013}, flux ratios of UV emission lines from BLRs in 70 PG QSOs 
at $z < 0.5$ were measured for studying $Z_{\rm BLR}$. Objects with 
broad absorption-line (BAL) features were excluded, for  
accurate measurements of the emission line fluxes. Among these 70 PG QSOs, 
31 objects have optical spectra in the data-archive of the Sloan Digital
Sky Survey \citep[SDSS;][]{York2000}. Since three objects show no strong 
NLR lines in their SDSS spectra, we select 28 PG QSOs with optical NLR spectra. 
Then we enlarge the sample size by examining the Markarian sample from the 
13th Veron-Cetty AGN catalog \cite{Veron2010}. Among 241 
Markarian AGNs, 30 objects have available UV and optical spectra. Through our 
visual inspection, we select 6 Markarian objects whose UV and optical spectra 
show high enough signal-to-noise ratios for our analysis  and show no BAL 
features. Therefore, we finalize sample that consists of 28 PG QSOs 
and 6 Markarian AGNs (34 non-BAL AGNs in total, see Table 1). 
The mean, standard deviation, and median of the redshift of 34 AGNs in our 
sample are 0.14, 0.11, and 0.13 respectively.

The UV spectra of our sample are retrieved from the Mikulski Archive for Space 
Telescopes (MAST). They were obtained with Cosmic Origins Spectrograph (COS), 
Space Telescope Imaging Spectrograph (STIS), or Faint Object Spectrograph 
(FOS) on board Hubble Space Telescope (HST), or the short-wavelength-prime 
(SWP) detector on board International Ultraviolet Explorer (IUE). If there are 
multiple data sets taken with different telescopes, we use higher resolution data 
in order of COS ($R \sim 16000-21000$), STIS ($R \sim 11400-17400$), FOS 
($R \sim 1300$), and IUE ($R \sim 300$). 
For some targets which were observed multiple times with the same instrument 
at similar epochs, we combine their spectra by calculating 
error-weighted mean to obtain their spectra with a better signal-to-noise ratio. 
For STIS and COS spectra, we perform 2 pixel boxcar smoothing for STIS spectra and 7 pixel boxcar smoothing for COS spectra 
since these spectra are highly over-sampled; i.e., the pixel scale is $\sim$ 0.05\AA/pixel for STIS spectra and $\sim$ 0.012\AA/pixel for COS spectra,
compared to the spectral resolution elements. 

For given a large difference of spectral resolution between HST and IUE observations,
we consider the effect of spectral resolution on the kinematic measurements from emission lines.
To examine whether our analyses are sensitive to the spectral resolution, we convolve two COS spectra (Mrk 106 and PG 1534+580) 
with a series of Gaussian kernels, to artificially decrease the spectral resolution. Using each of these spectra, we re0fit the emission lines.
We find that the outflow parameters (see \S3.1) and emission-line flux ratios used in 
our analyses are not sensitive to the spectral resolution, up to the spectral resolution $R \sim1000$. As expected
emission line flux ratios do not depend on the resolution and the kinematic measurements are consistent within $\sim$3-10\%.

The optical spectra of our sample are retrieved from the SDSS Data 
Release 7 \citep[DR7;][]{Abazajian2009}. The broad wavelength coverage of 
SDSS spectra (3800--9200\AA) enables us to detect various narrow emission 
lines (i.e., the [O~{\sc ii}] doublet emission at 3727.09\AA\ and 3729.88\AA,
the [O~{\sc iii}] doublet emission at 4960.30\AA\ and 5008.24\AA, and
the [S~{\sc ii}] doublet emission at 6718.32\AA\ and 6732.71\AA) 
for low-redshift AGNs. These narrow lines are used for the accurate redshift 
determination. Also, we can quantify various AGN properties such as the black 
hole mass and AGN bolometric luminosity using broad \Hb\ line and 5100\AA\ 
continuum luminosity. Details of our analysis on the SDSS optical spectra will be 
presented in \S 3.

Table 1 shows the details of the UV data (observing data, telescope, and 
instrument) for each object. Though the redshift given in Table 1 are taken from 
the NASA/IPAC Extragalactic Database (NED), we do not simply use those 
redshift for our analysis, (see \S 3.1).\\

%section 3
\section{Analysis}\label{section:analysis}
\subsection{Outflow indicators and redshift determination}\label{section: RD}

The BLR outflow is recognized through velocity profiles of broad emission lines in 
AGN spectra, characterized by the asymmetry or blueshift with respect to the 
systemic velocity of the object \citep[e.g.][]{Sulentic2000}. Each of the asymmetry 
and relative blueshift can be quantified straightforwardly based on observed 
emission-line spectra \citep[e.g.,][]{Sulentic2000, Baskin2005, Sulentic2007,
Wang2011}. \cite{Wang2011} proposed a new indicator, the blueshift 
and asymmetry index (BAI), that takes account of both the asymmetry and relative 
blueshift of emission lines \citep[see also][]{Wang2012}. In our analysis, 
we investigate 3 parameters to quantify the strength of the AGN outflow 
at the BLR; the asymmetry index, the velocity shift index, and BAI.

The asymmetric index (AI) is defined as the flux ratio between the blue part from the \CIV\ 
profile peak and the total (Eq. 1). The velocity shift index (VSI) is defined by the velocity
difference between the \CIV\ profile peak and the laboratory center of the \CIV\ 
emission, 1549.06\AA, which is the oscillator strength weighted average 
wavelength of the two \CIV\ doublet lines at 1548.20\AA\ and 1550.77\AA\
\citep[Eq. 2; see, e.g.,][]{Vandenberk2001}. The \CIV\ BAI, which combines the 
blueshift and asymmetric effects in the \CIV\ velocity profile, is the flux ratio 
between the blue part from the \CIV\ laboratory center and the total (Eq. 3). 
These indices are expressed by the following equations:

\begin{equation}              
% AI=\frac{Flux_{\rm Blue \ of \ peak}}{Flux_{\rm Total}}
AI = \frac{F\rm{(Blue \ region \ from \ the \ CIV \ peak)}}{F\rm{(Total \ CIV)}}
\label{eq:mbh}
\end{equation}
\begin{equation}        
% \CIV\ \rm velocity\ offset=  C~{\sc IV}_{\rm Peak} - C~{\sc IV}_{\rm Lab}
VSI = v{\rm (CIV \ peak)} - v{\rm (CIV \ lab)}
\label{eq:mbh}
\end{equation}
\begin{equation}        
%BAI=\frac{Flux_{\rm Blue \ of \ lab}}{Flux_{\rm Total}}
BAI = 
\frac{F{\rm(Blue \ region \ for \ the \ CIV \ laboratory \ wavelength)}}{F{\rm(Total \ CIV)}}
\label{eq:mbh}
\end{equation}

For determining VSI and BAI, the systemic redshift is crucial otherwise the laboratory location
of \CIV\ is not known. In principle, the best way to measure the systemic redshift of 
galaxies is to use stellar absorption-line features \citep[e.g.,][]{Bae2014,Woo2016}. Although 
measuring systematic redshift from stellar absorption-line is very 
challenging in type 1 AGNs due to strong AGN continuum emission 
\citep[but see, e.g.,][]{Woo2008, Harris2012, Park2012a, Park2015, Woo2015}, 
we try to measure stellar absorption lines (Mgb, Fe~{\sc ii}]$\lambda$5270, CaII 
triplet, and Ca H\&K lines) in SDSS spectra of our sample by using Penalized 
Pixel-Fitting method \citep[PPXF code][]{Cappellari2004}. This does not provide a good constraint, 
mainly due to insufficient signal-to-noise ratios of the SDSS 
spectra. Therefore we have to focus on some emission-line features, instead of 
stellar features, to determine the systemic redshift of each object.

Sometimes the UV \MgII\ emission line from the BLR is used to determine the
systemic redshift of type 1 AGNs \citep[e.g.,][]{Richards2002}, because the \MgII\ 
line is one of strong low-ionization emission lines from the BLR and thus less 
affected by nuclear radial motions of gas outflows as already mentioned in \S 1. 
However we do not use the \MgII\ line for determining the systemic redshift, 
because some of the UV spectra of our sample do not cover the \MgII\ line. 
Another reason is that the total mass of the 
BLR gas is only tiny with respect to the host galaxy \citep[e.g.,][]{Baldwin2003} 
and thus could be easily affected by possible radial motions. Some observations 
actually report that the \MgII-based redshift shows systematic differences from 
the systemic redshift (e.g., based on \OIII\ line; \citealt{McIntosh1999,
Vandenberk2001}).

More appropriate features than the \MgII\ line to determine the systemic redshift 
are narrow optical emission lines from NLRs, since the NLR resides at much large 
spatial scales \citep[e.g.,][]{Bennert2006b, Bennert2006a} with a larger total gas 
mass than the BLR \citep[e.g.,][]{Fraquelli2003, Crenshaw2015}. In the
previous studies, \OIII\ line is used to determine the systemic redshift 
\citep[e.g.,][]{HW10}.  However high-ionization NLR lines such as \OIII\ 
sometimes show asymmetric velocity profiles due to radial gas motions 
\citep[e.g.,][]{Komossa2008,Bae2014,Woo2015,Woo2016}, because such relatively high-ionization 
NLR lines arise at the innermost part of NLRs in AGNs (e.g., \citealt{Nagao2000,
Nagao2001a}) and thus easily affected by the outward pressure due to the AGN 
radiation. 
Therefore, in this work, we use low-ionization lines for determining the 
systemic redshift. Specifically, we use the \SII\, \OII, \OI, and 
H$\beta$ emission lines in this study. We fit those lines in the SDSS spectra by 
adopting a simple Gaussian profile. For the \SII\ (at 6718.29\AA\ and 
6732.67\AA\ in the laboratory) and \OI\ (at 6302.05\AA\ and 6365.5\AA) 
doublet emissions, we fix their wavelength separations 
and widths. We treated the \OII\ emission as a single Gaussian line at 
3728.48\AA\ (the average of the two doublet line wavelengths), because the 
\OII\ doublet lines are heavily blended in most cases. As for the \Hb\ and 
\OIII\ wavelength region, we subtract the blended Fe~{\sc II} multiplet and 
stellar emission before the narrow-line measurement. Note that the subtraction 
of the Fe~{\sc II} and stellar emission is important not only for the NLR fitting 
but also for measuring the velocity dispersion of the broad \Hb\ emission for 
the estimate of \mbh\ (see \S 3.2). We use the Fe~{\sc II} template given by 
\cite{Tsuzuki2006}, and the host galaxy template given by \cite{BC03},
a following procedures described in \citep{Park2012a,Park2015}.
After decomposing the broad component of \Hb\ with sixth-order of 
Gauss-Hermitian series, we fit the narrow component of \Hb\ and \OIII\ with 
the single Gaussian. We do not use \NII\ emission lines since it is buried by 
the broad \Ha\ line in most objects in our sample.

Using the measured central wavelength of each NLR line, we then calculate the 
systemic velocity (reference redshift), respectively, and estimate BAI and VSI, using the 
peak of \CIV\ line from the best-fit model.
To obtain the outflow and metallicity indicators, we fit emission lines in our interest,
namely, \CIV, \NV, \SiIV, \OIV,  and \HeII\ by adopting multi component fitting method \citep[see also][]{Shin2013}. 
First, we divide BLR lines into two groups based on the ionization potential, and assume 
that all emission lines in the same group have the same emission-line profile. 
Second, we mask out narrow absorption lines present in some targets. 
Third, we adopt a double Gaussian profile, which well reproduces 
the observed BLR emission lines. 
Here we assume that both Gaussian components are originated from the BLR
since the velocity width of the narrower Gaussian component is wider than 1500 km s$^{-1}$ in most cases,
while the contribution of the NLR emission to the CIV profile seems not significant as we 
do not detect them in our fitting analysis. Although it is possible that there is a weak narrow \CIV\ component,
the effect of this component on our results is presumably insignificant.
As we expect, high ionization emission lines (\NV, \CIV, \OIV, \HeII) are well fitted with the same line profile,
suggesting that other high ionization emission lines also show strong outflow as well as \CIV. 
However, \NV, \OIV, \HeII\ are blended with other emission lines. Thus, we use \CIV\ to avoid any systemic uncertainty due to the deblending.

In Figure 1, we compare the BAIs obtained with adopting different reference lines for a
consistency check. All VSIs measured 
based on each redshift reference line are correlated with each other with only small 
scatters, showing that there are no significant systematic difference among the VSIs 
derived from other emission lines. The scatter is mostly within a few tens km $\rm s^{-1}$ except for
the cases when the redshift given in the NED is adopted. In the following
analysis we exclude the reference redshift adopted from the NED. Instead, 
we decide to adopt the redshift reference in order of \SII, \OI, \OII, and \Hb, if not all lines are available. 
This order is determined based on the line 
strength and uncertainties. Note that the \Hb-based redshift potentially has a 
large uncertainty since the \Hb\ narrow component in type 1 AGNs is heavily blended 
with the \Hb\ broad component, and thus we adopt the \Hb-based redshift only 
when the redshift based on forbidden emission lines are unavailable. The 
measured BAI, VSI, and AI are given in Table 2, with the flag showing which 
narrow line is used as a reference for the systemic redshift.
For estimating the errors in BAI and VSI, we consider one resolution element of the SDSS 
spectrum (i.e., $\sim$70 km s$^{-1}$) as the 1 $\sigma$ uncertainty of the redshift, by assuming that the peak of the reference lines 
(e.g., \SII) would not move more than one pixel.
Regarding the uncertainty of AI, we simply calculate it by adding the blue-part flux error and red-part flux error, which are estimated based 
on the signal-to-noise ratio of each pixel. The estimated errors in BAI, VSI, and AI 
are also given in Table 2.\\

\begin{figure*}
\includegraphics[width = 0.96\textwidth]{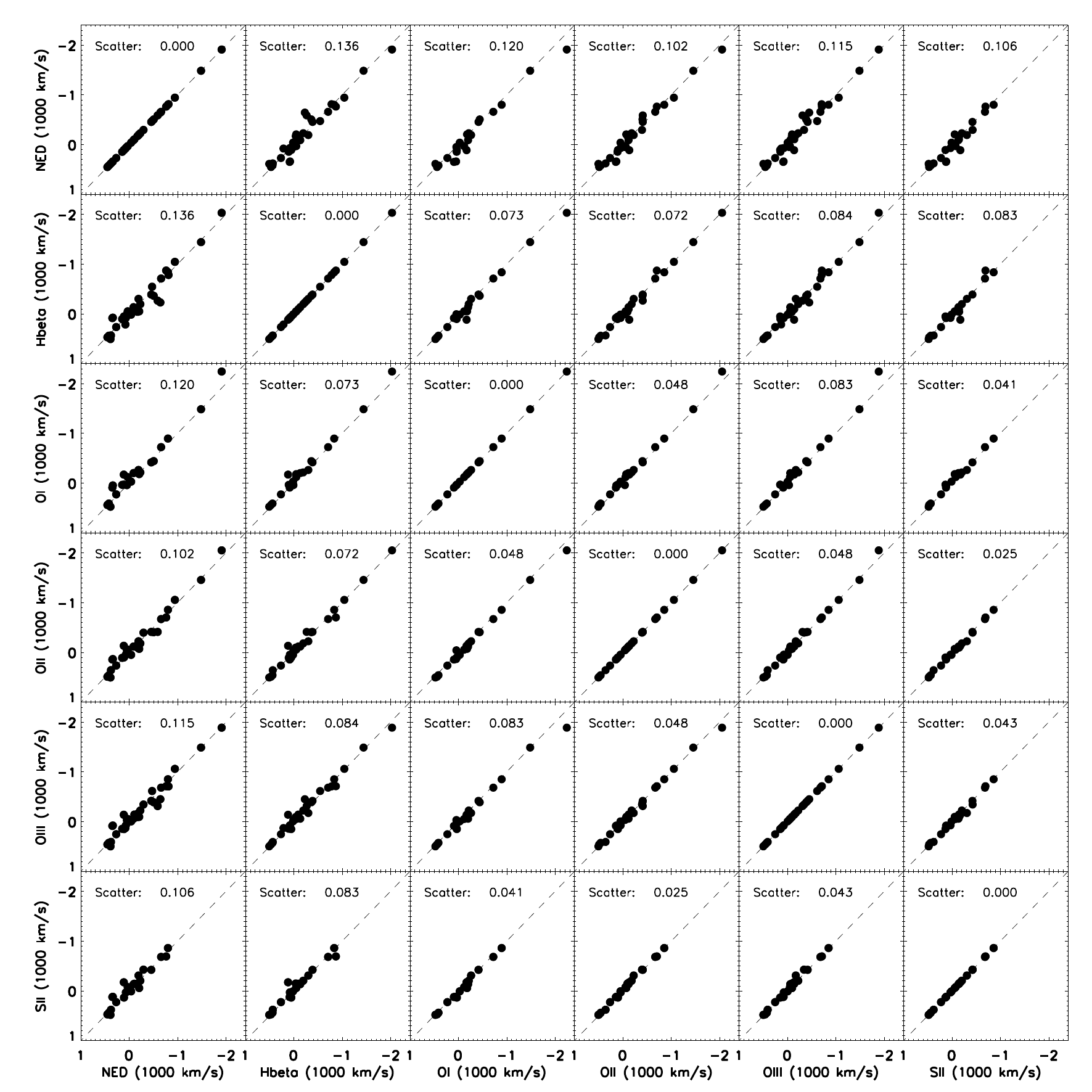}
\caption{
     The comparison of the VSIs derived by different redshifts determined
     by using different narrow lines or from the NED database. The scatter given
     in each panel is the standard deviation of the difference between the two 
     VSIs examined in each panel, in units of 1000 km $\rm s^{-1}$.
\label{fig:allspec1}}
\end{figure*}

\subsection{AGN properties} \label{section:Data}

Here we describe how we derive the AGN bolometric luminosity ($L_{\rm Bol}$),
\mbh, and Eddington ratio ($L_{\rm Bol}/L_{\rm Edd}$), that are compared with
the AGN outflow strength in \S 4. The AGN bolometric luminosity is derived from 
the monochromatic luminosity of the continuum emission at $\lambda_{\rm rest} = 
5100$\AA, where strong emission lines do not exist. 
We calculate the 5100A luminosity after subtracting the stellar component by adopting the 
template from \cite{BC03}, since in lower luminosity AGNs, the contribution of host galaxy emission is significant.
We adopt the bolometric correction factor of 9.0 to determine the AGN bolometric luminosity
\citep[e.g.,][]{Kaspi2000}. The broad component of the \Hb\ line seen in the 
SDSS spectrum is used to derive \mbh. Though sometimes \MgII\ and \CIV\ 
have been also used for estimating \mbh, we do not use them because our UV 
spectra are heterogeneous obtained with various instruments. 
Note that the \CIV\ emission line is widely used to infer the AGN outflow, 
implying that the \CIV\ velocity profile is largely affected by the AGN outflow 
\citep[i.e., the \CIV-emitting region may not be virialized;][]{Sulentic2007,Wang2011,Trakhtenbrot2012}. 
More specifically, we measure \mbh\ from the velocity dispersion of the \Hb\ 
broad component and the 5100\AA\ continuum luminosity by adopting the 
calibration given by \cite{Park2012b} and \cite{Woo2015}. The Eddington ratio is calculated simply 
from the derived AGN bolometric luminosity divided by the Eddington luminosity,
which is calculated using the measured \mbh. The derived \mbh\ and 
$L_{\rm Bol}/L_{\rm Edd}$ are given in Table 2, with the velocity dispersion of
the \Hb\ broad component and the 5100\AA\ continuum luminosity that are 
used for calculating \mbh\ and $L_{\rm Bol}/L_{\rm Edd}$.

The uncertainties in the \Hb\ velocity dispersion and \mbh\ are estimated by
performing Monte-Carlo simulations.  
 Specifically, we simulate 1000 mock spectra 
by randomizing flux with flux error \citep[see e.g.,][]{Bae2014,Woo2016}
and measure the velocity dispersion and \mbh\ from each mock spectrum. 
Then we take the standard deviations of the measurements as the uncertainties. Finally, the 
uncertainty in the Eddington ratio is estimated simply by combining the errors in 
the 5100\AA\ luminosity and \mbh. Table 2 shows the derived quantities with 
the estimated uncertainties, where systematic uncertainties are not included.

\subsection{AGN metallicity}\label{section:AGN}

The metallicity of gas clouds in the BLR has been often inferred from the flux 
ratios of \NV/\CIV\ and \NV/\HeII, since the nitrogen relative abundance scales
to the square of the metallicity (i.e., N/H $\propto$ $Z_{\rm BLR}^2$, or 
equivalently, N/O $\propto$ O/H $\propto$ $Z_{\rm BLR}$) due to the nature of
the nitrogen as a secondary element (see \citealt{Hamann1999} and references 
therein). Through intensive calculations of photoionization models,
\cite{Nagao2006} proposed an alternative flux ratio as an indicator of 
$Z_{\rm BLR}$, (\SiIV+\OIV)/\CIV, that has been also used to infer 
$Z_{\rm BLR}$ especially when the deblending of the \NV\ emission from the
neighboring Ly$\alpha$ emission is difficult \citep[e.g.,][]{Juarez2009,
Matsuoka2011b,Wang2012}.
In this study, we examine the three indicators for $Z_{\rm BLR}$, that are 
\NV/\CIV, \NV/\HeII, and (\SiIV+\OIV)/\CIV. These flux ratios of PG QSOs have 
been already measured by \cite{Shin2013}, taking the careful decomposition of 
blended lines such as \NV. We adopt the same measurement method also for the 
Markarian sample in this study. We present the measured flux and its uncertainty of 
the emission lines used for inferring $Z_{\rm BLR}$ in Table 2. In this table we do 
not include the flux of \HeII\ and \SiIV+\OIV\ for some objects, because the flux 
cannot be measured accurately for those cases due to their low signal-to-noise 
ratio.\\

\begin{turnpage}
\begin{deluxetable*}{lcrccccccrrrr} 
\tablewidth{0pt}
\tablecolumns{10}
\tabletypesize{\scriptsize}
\tablecaption{Outflow indices, AGN properties, and UV emission-line fluxes}
\tablehead{
\colhead{Object} &
\colhead{AI} &
\colhead{VSI} &
\colhead{BAI} &
\colhead{Flag} &
\colhead{$\rm \sigma_{H\beta}$} &
\colhead{log[$\lambda L_{5100}$]}&
\colhead{log[$M_{\rm BH}/M_{{\odot}}$]} &
\colhead{log[$L_{\rm Bol}/L_{\rm Edd}$]} &
\colhead{N V} &
\colhead{Si IV+O IV]} &
\colhead{C IV}&
\colhead{He II}
\\
\colhead{} &
\colhead{} &
\colhead{(km s$^{-1}$)} &
\colhead{} &
\colhead{} &
\colhead{(km s$^{-1}$)} &
\colhead{($\rm erg \rm\ s^{- 1}$)}&
\colhead{}&
\colhead{}&
\colhead{\tiny(10$^{-14}$ \ergs\ $\rm cm^{-2}$)} &
\colhead{\tiny(10$^{-14}$ \ergs\ $\rm cm^{-2}$)} &
\colhead{\tiny(10$^{-14}$ \ergs\ $\rm cm^{-2}$)} &
\colhead{\tiny(10$^{-14}$ \ergs\ $\rm cm^{-2}$)} 
\\
\colhead{(1)} &
\colhead{(2)} &
\colhead{(3)} &
\colhead{(4)} &
\colhead{(5)}&
\colhead{(6)} &
\colhead{(7)} &
\colhead{(8)} &
\colhead{(9)} &
\colhead{(10)} &
\colhead{(11)} &
\colhead{(12)} &
\colhead{(13)} 
}
\startdata             
Mrk~106		&$	0.59	\pm	0.05	$&$	-76	\pm	60	$&$	0.60	\pm	0.02	$&$	1	$&$	2439	\pm	26	$&$	44.45	\pm	0.01	$&$	8.46	\pm	0.01	$&$	-1.15	\pm	0.01	$&$		26.8	\pm	2.1	$&$	8.7	\pm	0.9		$&$	54.2	\pm	9.0		$&$	{--}				$\\
Mrk~110		&$	0.54	\pm	0.04	$&$	472	\pm	65	$&$	0.43	\pm	0.02	$&$	1	$&$	1992	\pm	48	$&$	42.87	\pm	0.01	$&$	7.45	\pm	0.02	$&$	-1.72	\pm	0.02	$&$		54.5	\pm	5.4	$&$	47.2	\pm	5.2		$&$	227.2	\pm	20.3		$&$	22.2 \pm 1.4				$\\
Mrk~290		&$	0.55	\pm	0.03	$&$	119	\pm	66	$&$	0.46	\pm	0.01	$&$	1	$&$	2683	\pm	32	$&$	43.42	\pm	0.01	$&$	8.02	\pm	0.01	$&$	-1.75	\pm	0.01	$&$		45.0	\pm	4.6	$&$	22.1	\pm	1.4		$&$	172.0	\pm	20.7		$&$	15.1\pm 2.4				$\\
Mrk~493		&$	0.49	\pm	0.07	$&$	-133	\pm	66	$&$	0.55	\pm	0.02	$&$	1	$&$	887	\pm	175	$&$	43.19	\pm	0.01	$&$	6.83	\pm	0.17	$&$	-0.79	\pm	0.17	$&$		28.0	\pm	6.5	$&$	18.3	\pm	2.8		$&$	56.4	\pm	13.2		$&$	7.1\pm 1.1				$\\
Mrk~506		&$	0.48	\pm	0.05	$&$	227	\pm	65	$&$	0.45	\pm	0.01	$&$	1	$&$	3078	\pm	150	$&$	42.85	\pm	0.01	$&$	7.85	\pm	0.05	$&$	-2.14	\pm	0.05	$&$		62.7	\pm	7.3	$&$	34.5	\pm	4.1		$&$	218.2	\pm	23.7		$&$	17.5\pm 1.8				$\\
Mrk~1392		&$	0.57	\pm	0.04	$&$	9	\pm	66	$&$	0.58	\pm	0.01	$&$	1	$&$	2859	\pm	32	$&$	43.32	\pm	0.01	$&$	8.03	\pm	0.01	$&$	-1.85	\pm	0.01	$&$		11.4	\pm	0.6	$&$	9.7	\pm	0.5		$&$	87.4	\pm	8.4		$&$	6.6\pm 1.0				$\\
PG~0157+001		&$	0.55	\pm	0.04	$&$	-1483	\pm	63	$&$	0.82	\pm	0.01	$&$	2	$&$	1515	\pm	95	$&$	44.71	\pm	0.01	$&$	8.14	\pm	0.06	$&$	-0.57	\pm	0.06	$&$		29.2	\pm	2.1	$&$	{--}				$&$	64.0	\pm	2.8		$&$	3.8	\pm	0.2		$\\
PG~0921+525		&$	0.52	\pm	0.03	$&$	440	\pm	65	$&$	0.36	\pm	0.02	$&$	1	$&$	1934	\pm	49	$&$	42.87	\pm	0.01	$&$	7.42	\pm	0.03	$&$	-1.69	\pm	0.03	$&$		39.7	\pm	1.9	$&$	46.9	\pm	2.8		$&$	292.0	\pm	12.9		$&$	13.7	\pm	0.5		$\\
PG~0923+129		&$	0.53	\pm	0.07	$&$	482	\pm	66	$&$	0.45	\pm	0.01	$&$	1	$&$	1679	\pm	53	$&$	43.11	\pm	0.01	$&$	7.41	\pm	0.03	$&$	-1.44	\pm	0.03	$&$		66.3	\pm	5.0	$&$	{--}				$&$	183.3	\pm	15.8		$&$	18.4	\pm	1.3		$\\
PG~0947+396		&$	0.51	\pm	0.03	$&$	-143	\pm	56	$&$	0.55	\pm	0.01	$&$	1	$&$	2804	\pm	27	$&$	44.65	\pm	0.01	$&$	8.70	\pm	0.01	$&$	-1.19	\pm	0.02	$&$		32.9	\pm	2.0	$&$	11.5	\pm	0.7		$&$	59.5	\pm	2.6		$&$	7.3	\pm	0.3		$\\
PG~1022+519		&$	0.49	\pm	0.07	$&$	31	\pm	65	$&$	0.51	\pm	0.02	$&$	1	$&$	965	\pm	144	$&$	43.35	\pm	0.01	$&$	7.00	\pm	0.13	$&$	-0.79	\pm	0.13	$&$		12.2	\pm	1.1	$&$	{--}				$&$	45.7	\pm	4.5		$&$	{--}				$\\
PG~1048+342		&$	0.51	\pm	0.13	$&$	-147	\pm	58	$&$	0.54	\pm	0.01	$&$	1	$&$	2328	\pm	43	$&$	44.42	\pm	0.01	$&$	8.40	\pm	0.02	$&$	-1.13	\pm	0.02	$&$		12.4	\pm	2.9	$&$	{--}				$&$	19.1	\pm	3.4		$&$	{--}				$\\
PG~1049-005		&$	0.52	\pm	0.12	$&$	-681	\pm	50	$&$	0.68	\pm	0.01	$&$	1	$&$	2925	\pm	21	$&$	45.51	\pm	0.01	$&$	9.19	\pm	0.01	$&$	-0.82	\pm	0.01	$&$		29.7	\pm	7.3	$&$	11.3	\pm	1.9		$&$	41.6	\pm	6.4		$&$	4.7	\pm	0.7		$\\
PG~1115+407		&$	0.50	\pm	0.04	$&$	-690	\pm	59	$&$	0.63	\pm	0.01	$&$	1	$&$	1830	\pm	73	$&$	44.59	\pm	0.01	$&$	8.26	\pm	0.04	$&$	-0.81	\pm	0.04	$&$		18.5	\pm	1.2	$&$	{--}				$&$	34.4	\pm	2.0		$&$	{--}				$\\
PG~1121+422		&$	0.47	\pm	0.05	$&$	213	\pm	76	$&$	0.40	\pm	0.02	$&$	4	$&$	1834	\pm	45	$&$	44.94	\pm	0.01	$&$	8.44	\pm	0.02	$&$	-0.64	\pm	0.02	$&$		14.5	\pm	1.9	$&$	{--}				$&$	47.8	\pm	3.5		$&$	{--}				$\\
PG~1151+117		&$	0.51	\pm	0.05	$&$	-229	\pm	80	$&$	0.55	\pm	0.01	$&$	4	$&$	2734	\pm	31	$&$	44.67	\pm	0.01	$&$	8.68	\pm	0.01	$&$	-1.16	\pm	0.01	$&$		35.9	\pm	3.9	$&$	{--}				$&$	51.1	\pm	3.5		$&$	6.6	\pm	0.5		$\\
PG~1202+281		&$	0.56	\pm	0.05	$&$	-860	\pm	58	$&$	0.76	\pm	0.01	$&$	1	$&$	3183	\pm	34	$&$	44.12	\pm	0.01	$&$	8.54	\pm	0.01	$&$	-1.57	\pm	0.02	$&$		8.3	\pm	0.8	$&$	{--}				$&$	72.9	\pm	5.1		$&$	{--}				$\\
PG~1229+204		&$	0.50	\pm	0.02	$&$	-172	\pm	64	$&$	0.50	\pm	0.01	$&$	1	$&$	2384	\pm	29	$&$	43.70	\pm	0.01	$&$	8.05	\pm	0.01	$&$	-1.49	\pm	0.01	$&$		15.0	\pm	0.6	$&$	47.5	\pm	2.0		$&$	156.4	\pm	4.8		$&$	11.4	\pm	0.3		$\\
PG~1244+026		&$	0.50	\pm	0.08	$&$	-309	\pm	65	$&$	0.67	\pm	0.02	$&$	1	$&$	853	\pm	267	$&$	43.41	\pm	0.01	$&$	6.91	\pm	0.28	$&$	-0.64	\pm	0.28	$&$		3.7	\pm	0.3	$&$	{--}				$&$	11.4	\pm	1.2		$&$	{--}				$\\
PG~1307+085		&$	0.54	\pm	0.03	$&$	-55	\pm	59	$&$	0.55	\pm	0.01	$&$	1	$&$	2459	\pm	25	$&$	44.74	\pm	0.01	$&$	8.62	\pm	0.01	$&$	-1.02	\pm	0.01	$&$		64.3	\pm	4.2	$&$	{--}				$&$	114.6	\pm	4.6		$&$	{--}				$\\
PG~1341+258		&$	0.50	\pm	0.12	$&$	379	\pm	62	$&$	0.42	\pm	0.01	$&$	1	$&$	1824	\pm	58	$&$	43.75	\pm	0.01	$&$	7.82	\pm	0.03	$&$	-1.21	\pm	0.03	$&$		22.4	\pm	2.6	$&$	{--}				$&$	34.7	\pm	5.3		$&$	{--}				$\\
PG~1404+226		&$	0.51	\pm	0.09	$&$	-2240	\pm	67	$&$	0.91	\pm	0.01	$&$	2	$&$	1205	\pm	110	$&$	44.13	\pm	0.01	$&$	7.61	\pm	0.09	$&$	-0.63	\pm	0.09	$&$		9.5	\pm	1.3	$&$	{--}				$&$	13.1	\pm	1.6		$&$	{--}				$\\
PG~1415+451		&$	0.57	\pm	0.04	$&$	-425	\pm	61	$&$	0.65	\pm	0.01	$&$	1	$&$	1713	\pm	77	$&$	43.95	\pm	0.01	$&$	7.86	\pm	0.04	$&$	-1.05	\pm	0.04	$&$		44.0	\pm	2.7	$&$	20.4	\pm	1.1		$&$	55.0	\pm	2.8		$&$	6.6	\pm	0.3		$\\
PG~1425+267		&$	0.51	\pm	0.02	$&$	-442	\pm	53	$&$	0.55	\pm	0.01	$&$	2	$&$	4068	\pm	19	$&$	45.14	\pm	0.01	$&$	9.31	\pm	0.00	$&$	-1.32	\pm	0.01	$&$		8.6	\pm	0.4	$&$	{--}				$&$	49.4	\pm	1.5		$&$	{--}				$\\
PG~1427+480		&$	0.48	\pm	0.03	$&$	136	\pm	55	$&$	0.44	\pm	0.01	$&$	1	$&$	2019	\pm	39	$&$	44.56	\pm	0.01	$&$	8.34	\pm	0.02	$&$	-0.92	\pm	0.02	$&$		13.8	\pm	0.7	$&$	9.3	\pm	0.4		$&$	43.1	\pm	2.0		$&$	6.1	\pm	0.3		$\\
PG~1444+407		&$	0.50	\pm	0.06	$&$	-780	\pm	74	$&$	0.65	\pm	0.01	$&$	4	$&$	2447	\pm	31	$&$	45.14	\pm	0.01	$&$	8.82	\pm	0.01	$&$	-0.82	\pm	0.01	$&$		36.0	\pm	2.1	$&$	{--}				$&$	32.6	\pm	2.8		$&$	{--}				$\\
PG~1448+273		&$	0.56	\pm	0.10	$&$	-421	\pm	64	$&$	0.72	\pm	0.02	$&$	1	$&$	1209	\pm	171	$&$	44.18	\pm	0.01	$&$	7.65	\pm	0.13	$&$	-0.61	\pm	0.13	$&$		10.0	\pm	0.8	$&$	3.7	\pm	0.2		$&$	13.2	\pm	1.5		$&$	4.1	\pm	0.5		$\\
PG~1512+370		&$	0.57	\pm	0.08	$&$	41	\pm	53	$&$	0.56	\pm	0.01	$&$	2	$&$	3447	\pm	28	$&$	45.18	\pm	0.01	$&$	9.17	\pm	0.01	$&$	-1.13	\pm	0.02	$&$		32.9	\pm	5.8	$&$	8.3	\pm	1.0		$&$	70.2	\pm	7.3		$&$	4.5	\pm	0.8		$\\
PG~1519+226		&$	0.44	\pm	0.08	$&$	-411	\pm	108	$&$	0.53	\pm	0.02	$&$	3	$&$	1552	\pm	45	$&$	44.34	\pm	0.01	$&$	7.97	\pm	0.03	$&$	-0.77	\pm	0.03	$&$		22.4	\pm	3.4	$&$	29.0	\pm	6.0		$&$	49.0	\pm	5.9		$&$	{--}				$\\
PG1534+580		&$	0.54	\pm	0.02	$&$	124	\pm	66	$&$	0.47	\pm	0.01	$&$	1	$&$	2701	\pm	30	$&$	43.42	\pm	0.01	$&$	8.03	\pm	0.01	$&$	-1.75	\pm	0.01	$&$		40.8	\pm	1.5	$&$	25.1	\pm	0.5		$&$	187.5	\pm	6.0		$&$	14.3	\pm	0.6		$\\
PG~1545+210		&$	0.49	\pm	0.04	$&$	34	\pm	57	$&$	0.49	\pm	0.01	$&$	2	$&$	3248	\pm	20	$&$	45.15	\pm	0.01	$&$	9.10	\pm	0.01	$&$	-1.09	\pm	0.01	$&$		36.9	\pm	2.0	$&$	{--}				$&$	101.6	\pm	5.4		$&$	6.9	\pm	0.4		$\\
PG~1552+085		&$	0.48	\pm	0.10	$&$	-545	\pm	84	$&$	0.60	\pm	0.02	$&$	4	$&$	1245	\pm	95	$&$	44.30	\pm	0.01	$&$	7.73	\pm	0.07	$&$	-0.57	\pm	0.07	$&$		12.7	\pm	2.2	$&$	{--}				$&$	30.7	\pm	4.4		$&$	{--}				$\\
PG~1612+261		&$	0.50	\pm	0.06	$&$	-203	\pm	60	$&$	0.54	\pm	0.01	$&$	1	$&$	2108	\pm	45	$&$	44.41	\pm	0.01	$&$	8.30	\pm	0.02	$&$	-1.04	\pm	0.02	$&$		11.0	\pm	1.3	$&$	{--}				$&$	67.8	\pm	5.1		$&$	2.5	\pm	0.1		$\\
PG~2233+134		&$	0.53	\pm	0.02	$&$	-1052	\pm	93	$&$	0.74	\pm	0.02	$&$	3	$&$	2224	\pm	64	$&$	45.12	\pm	0.01	$&$	8.72	\pm	0.03	$&$	-0.74	\pm	0.03	$&$		10.4	\pm	0.8	$&$	4.4	\pm	0.2		$&$	8.0	\pm	0.3		$&$	{--}				$
\enddata
\label{table:prop}
\tablecomments{
     Col. (5): Flag for reference lines used to determine the redshift. 
     1: \SII, 2: \OI, 3: \OII, and 4: \Hb\ narrow component. \\
}
\end{deluxetable*}
\end{turnpage}

%section 4
\section{result} \label{section:result}

%We present the result of comparison between outflow indicators (AI, \CIV\ 
%velocity offset, and BAI). Then, we compare outflow indicators with other 
%AGN parameters, (black hole mass, AGN luminosity, Eddington ratio, and 
%AGN metallicity).

\begin{figure*}{}
\begin{center}
\includegraphics[width = 0.76\textwidth]{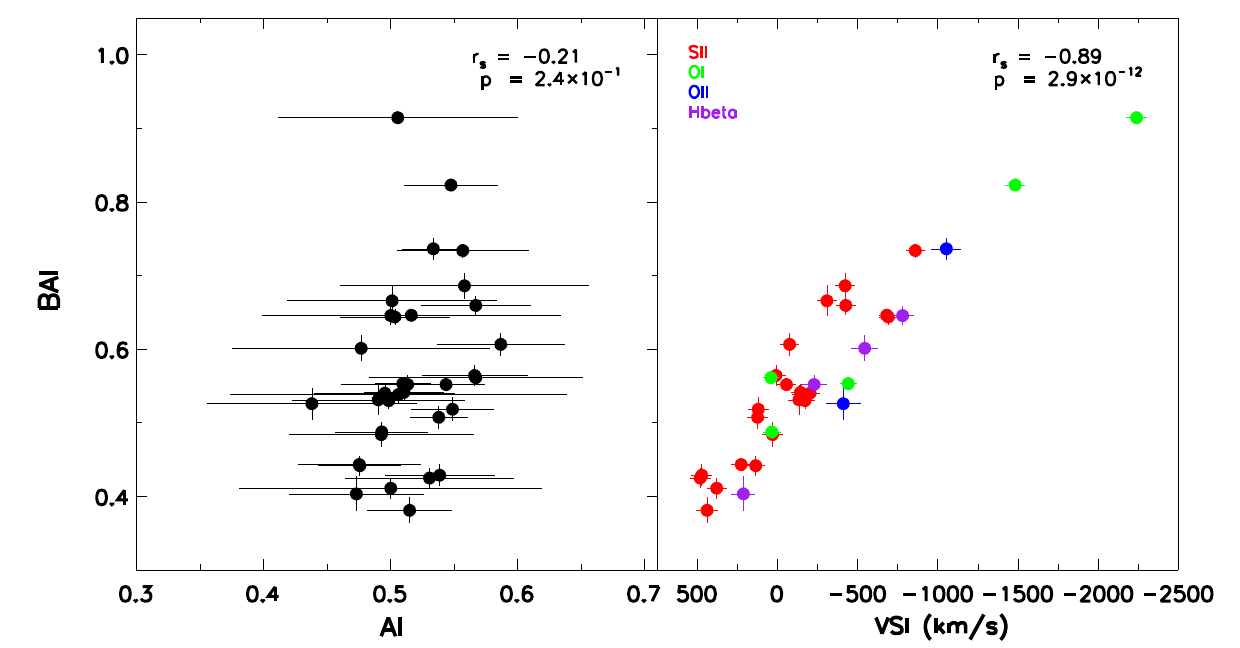}
\end{center}
\caption{
     Relation among the outflow indicators. Left panel: comparison of AI and BAI. 
     Right panel: comparison of VSI versus BAI. The colors in right panel represent 
     the reference lines for determining the systemic redshift (\SII: red, \OI: green, 
     \OII: blue, and H$\beta$: purple). Spearman's rank correlation coefficients and 
     their statistical significances are given in the top left side in each panel.}
\label{fig:allspec1}
\end{figure*}

%section 4.1
\subsection{Outflow indicators}

As described in \S3.1, the outflow indicators investigated in this work are AI, 
VSI, and BAI. The mean values and standard deviations for AI, VSI, and BAI 
are (0.51, 0.03), (--258 km s$^{-1}$, 566 km s$^{-1}$), and (0.57, 0.12), 
respectively. For comparison, the median value of those parameters given in 
\cite{Wang2011} for high-redshift QSOs are 0.5 (AI), -898 km s$^{-1}$ (VSI), 
and 0.63 (BAI). While the measured AI and BAI of our sample are similar to 
those of the high-redshift QSO sample of \cite{Wang2011}, the VSI values of 
our sample are systematically smaller than those of \cite{Wang2011}. The 
reason for this systematic difference may be owing to the difference in the 
sample selection (see \S5.1).
% The difference in $L_{\rm Bol}/L_{\rm Edd}$ between the two samples will be
% checked in \S 4.2.

To examine the reliability of the measured BAI and the relation between BAI and 
the other indicators (AI and VSI), we compare the three outflow indicators in 
Figure 2. The different colors of symbols in this figure denote different narrow lines
used for determining the systemic redshift. We do not find any systematically
different behaviors for different color symbols, suggesting that the difference in
the narrow lines used for the redshift determination does not cause significant
systematic effects in the outflow indicators. Therefore we combine all of the data
obtained from different reference lines in the following analysis and discussion.
Figure 2 shows that there is a clear correlation between BAI and VSI, while there
is no apparent correlation between BAI and AI. To investigate whether or not 
there are statistically significant correlations among the outflow indices, we 
conduct Spearman's rank-order correlation test with the null hypothesis that there
are no correlations among the outflow indices. This null hypothesis is not rejected
for the relation between BAI and AI ($p = 2.4 \times 10^{-1}$), while it is rejected
for the relation between BAI and VSI with a high statistical significance ($p =
2.9 \times 10^{-12}$). This result is partly because the typical measurement 
uncertainty in AI is much larger than the standard deviation of the AI distribution
(see the left panel in Figure 2). It is thus inferred that the AI parameter is not an
adequate indicator for the AGN outflow, at least for our sample. 
Note that BAI and VSI are expected to show a correlation since both parameters
are defined with velocity shift from the systematic velocity. Thus, it is not surprising
to observe a correlation in Figure 2 (left).\\

\subsection{AGN properties and outflow}\label{section: comparison}

Here we investigate the relation between outflow indicators (BAI and VSI) and 
AGN properties. The mean values and standard deviations of 
 $M_{\rm BH}/M_{{\odot}}$, $\lambda L_{5100} \ (\rm erg/s)$, and
$L_{\rm Bol}/L_{\rm Edd}$ derived in \S 3.2 in logscale, 
are (8.20, 0.61), (45.23, 0.70), and (--1.07, 0.41), respectively. The bolometric
luminosity of our sample is located at the bright part of the optical AGN luminosity
function in the local universe \citep[see, e.g.,][]{Boyle2000, Schulze2009}.
As for the Eddington ratio, our
sample shows a wide range ($\sim$1.5 dex) in the Eddington ratio, distributed in
the range of $-2.0 \lesssim L_{\rm Bol}/L_{\rm Edd} \lesssim -0.5$.
Thus our sample allows us to investigate the AGN outflow in a broad dynamic 
range of the Eddington ratio. Note that the Eddington ratio of the sample of
\cite{Wang2011} distributes in the range of $-1.0 \lesssim 
L_{\rm Bol}/L_{\rm Edd} \lesssim +0.5$, i.e., $\sim$1 dex higher than the
Eddington ratio distribution of our sample. This difference is a natural 
consequence of the sample selection in the sense that high-redshift SDSS
quasars are luminous enough to be selected in the magnitude-limited
SDSS spectroscopy. The systematically smaller VSI in our sample than in the
sample of \cite{Wang2011} mentioned in \S4.1 is probably attributed by this
$\sim$1 dex difference in the Eddington ratio distribution between the two
samples.

In Figure 3, possible dependences of the two outflow indicators (BAI and VSI) on
the AGN properties ($M_{\rm BH}$, $L_{\rm Bol}$, and $L_{\rm Bol}/L_{\rm Edd}$)
are investigated. Note that the investigation of the relation between AI and those
AGN properties is not useful, since the dynamic range of AI in our sample is too
narrow (see \S4.1 and Figure 2). Figure 3 shows that there is no apparent 
correlation between \mbh\ and outflow indicators. However in the center and right 
panel, the two outflow indicators appear to show positive correlations with the 
AGN bolometric luminosity and Eddington ratio, in the sense that AGNs with a 
higher luminosity or a higher Eddington ratio tend to show stronger outflows. For 
examining the statistical significance of these possible correlations, we conduct 
Spearman's rank-order correlation test with the null hypothesis that there are no 
correlations in Figure 3. The rank-order test suggests that there is no correlation
between the two outflow indicators (BAI and VSI) and \mbh\ ($p = 1.6 \times 
10^{-1}$ and $p = 1.3 \times 10^{-1}$). It is also suggested that the inferred 
dependences of BAI and VSI on the AGN bolometric luminosity and Eddington 
ratio are marginal, but VSI shows more significant correlation with those two
AGN parameters ($p = 4.7 \times 10^{-3}$ and $p = 1.1 \times 10^{-3}$) than BAI 
($p = 1.9 \times 10^{-2}$ and $p = 1.1 \times 10^{-2}$). These trends are 
qualitatively consistent with the previous study for high-redshift QSOs reported 
by \cite{Wang2011}.

\begin{figure*}{}
\begin{center}
\includegraphics[width = 0.86\textwidth]{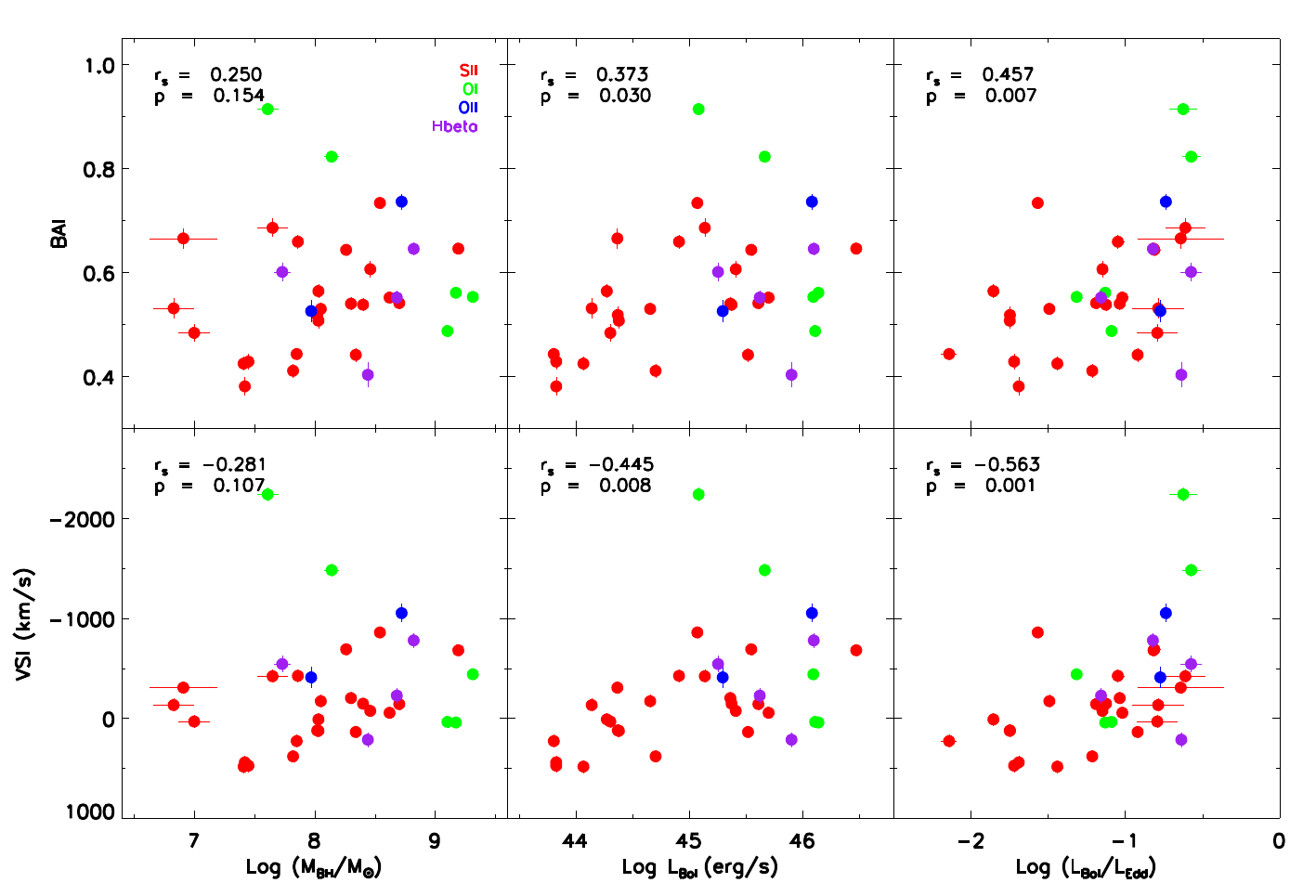}
\end{center}
\caption{
     Comparison between the AGN properties ($M_{\rm BH}$, $L_{\rm bol}$, 
     and $L_{\rm bol}/L_{\rm Edd}$) and the outflow indicators (BAI and VSI). 
     The colors are the same as in Figure 2. Spearman's rank correlation 
     coefficients and their statistical significance are presented at the top-left side 
     in each panel.
\label{fig:allspec1}}
\end{figure*}

\subsection{Comparison between metallicity and outflow}\label{section:comparison}

The mean values and standard deviations of \NV/\CIV, (\SiIV+\OIV)/\CIV, and 
\NV/\HeII derived in \S 3.3 in the logscale are (--0.41,0.29), (--0.65, 0.23), 
and (0.57, 0.22), respectively. As we mentioned in \S 3.3 the values are from 
\cite{Shin2013} and they are similar to those of previous works 
\cite[e.g.,][]{Shemmer2004, Matsuoka2011b}. 

Now we examine how the two outflow indicators (BAI and VSI) depend on the BLR
metallicity. Figure 4 shows the comparison between the outflow indicators and BLR
metallicity indicators. There are loose positive correlations in the sense that AGNs
with a higher BLR metallicity tend to show stronger outflow, but the significance 
looks marginal. For examining the statistical significance of those possible 
correlations, we conduct Spearman's rank-order correlation test with the null 
hypothesis that there are no correlations in Figure 4. The rank-order test suggests 
that the correlations between BAI and the metallicity indicators are very marginal 
with low statistical significance ($p \sim 0.01 - 0.17$). Though the statistical 
significances of correlations between VSI and the metallicity indicators are slightly
higher ($p \sim 0.003 - 0.07$) than the correlations for BAI, but not very tight.
These results seem contradictory to the earlier results for high redshift QSOs 
reported by \cite{Wang2012}, who showed a tight correlation between the 
outflow strength and metallicity in BLRs. For this comparison, it should be noted
that the analysis by \cite{Wang2012} is based on stacked SDSS spectra of
12844 high-redshift QSOs while our analysis is based on individual SDSS spectra 
of 34 low-redshift AGNs. A significantly larger number of objects in the study of 
\cite{Wang2012} makes the statistical error very small. 

For investigating whether
our results are consistent with those of \cite{Wang2012}, we present the relation between BAI
and (\SiIV+\OIV)/\CIV\ high-redshift QSOs in Figure 4 (dashed line in the upper middle panel).
 Although the dispersion in our sample is large,
the results of \cite{Wang2012} and ours show similar trend. This is interesting since the 
difference in the cosmic age between our sample at $z \sim 0.1$ and the sample 
of \cite{Wang2012} at $1.7<z<4$ is more than half of the Hubble time but still 
these two samples show a similar trend in the relation between the outflow 
indicator and metallicity indicator.

\begin{figure*}{}
\begin{center}
\includegraphics[width = 0.86\textwidth]{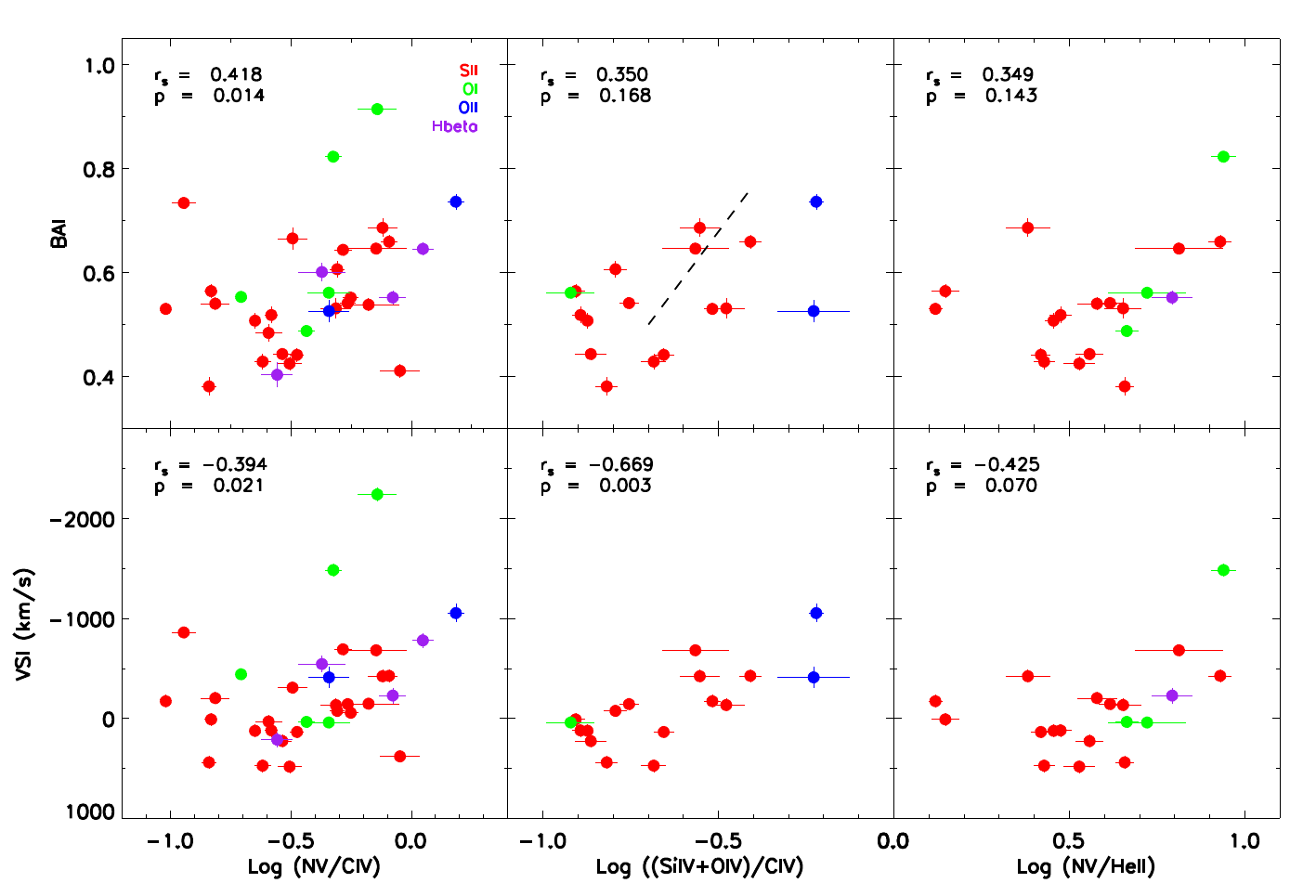}
\end{center}
\caption{
     Comparison between the metallicity indicators (\NV/\CIV, (\SiIV+\OIV)/\CIV, 
     and \NV/\HeII) and the outflow indicators (BAI and VSI). 
     The colors are the same as in Figure 2. Spearman's rank correlation 
     coefficients and their statistical significance are presented at the top-left side 
     in each panel. The dashed line in the top-middle panel shows the previous
     result for high-redshift QSOs reported by \cite{Wang2012}.
\label{fig:allspec1}}
\end{figure*}

%section 5
\section{Discussion}\label{section:Discussion}

%\subsection{Reliability of AGN outflow indicators}\label{OIII}
\subsection{Possible effects of the sample selection}

As described in \S4.2, the AI parameter does not work well for quantifying the 
outflow strength for our sample. This is attributed to the narrow range of the AI 
distribution. It is interesting to notice that highly asymmetric \CIV\ lines are 
selectively seen in radio-loud QSOs \citep{Sulentic2000, Sulentic2007}. The 
small range of the AI distribution in our sample may be attributed to the fact that 
there are not so many radio-loud AGNs in the PG sample \citep{Kellermann1989}.
\cite{Wang2011} reported a significant correlation between AI and BAI, based on
their high-redshift SDSS QSO sample whose AI distributes in the range of
$0.3 \lesssim AI \lesssim 0.8$. However, the AI distribution of their sample 
concentrates mostly in the range of $0.4 \lesssim AI \lesssim 0.6$, and actually
the wide AI range is achieved owing to the large sample size. Therefore, the 
small sample size of ours (lacking radio-loud AGNs especially) is not adequate 
to investigate the AI distribution and various AI-related correlations. 
As for the BAI and VSI parameters, VSI shows tighter correlations with AGN
properties including the BLR metallicity than BAI as shown in \S4.2 and 
\S4.3. This is possibly affected by the large uncertainty in AI that introduce the 
corresponding uncertainty in BAI.

Another potential concern for our sample selection is the removal of BAL AGNs.
The BAL features are caused by powerful outflows of ionized gas, and thus 
possibly AGNs with a strong outflow may be selectively removed from our sample.
This concern may not be true, if the fast ionized outflow is ubiquitous in AGNs and
the BAL features are observed just due to the orientation effect (i.e., we see the
BAL features only in $\sim$10\% of type-1 AGNs because the covering factor of
the ionized outflow is $\sim$10\%; see, e.g., \citealt{Weymann1991}). It is interesting
to compare the outflow indices (AI, VSI, and BAI) between BAL AGNs and non-BAL
AGNs to understand the nature of the BAL phenomenon in AGNs; however we do 
not further discuss this issue because it is beyond the scope of this work.\\

\subsection{Star formation and AGN activity}\label{starformation}

The correlations among the outflow, metallicity, and Eddington ratio seen in our
sample suggest their physical connection in low-redshift AGNs. It suggests that the 
gas radial motion (i.e., the gas accretion onto the SMBH and the outflow from the 
nucleus) is related to the metallicity in the nuclear region. The relation between the 
metallicity and the gas accretion has been reported for QSOs 
\citep[e.g.,][]{Shemmer2004}, and also for nearby Seyfert galaxies especially 
focusing on the metallicity difference between narrow-line Seyfert 1 galaxies and 
broad-line Seyfert 1 galaxies \citep[e.g.,][]{Nagao2002, Shemmer2002, Fields2005}.
This can be interpreted as the result of the starburst-AGN connection; i.e., the 
nuclear starburst activity enriches the gas in the nuclear region and also triggers 
the gas accretion onto the SMBH. 
On the other hand, the relation between the AGN outflow and the Eddington ratio has
been also discussed, in the sense that the gas accretion results in the increase of the
AGN luminosity that causes the outward radiative pressure to the surrounding gas
\citep[e.g.,][]{Komossa2008, Wang2011, Marziani2013}. Therefore the correlations 
among the outflow, metallicity, and Eddington ratio are reasonable characteristics in 
AGNs. In addition, a more direct reason causing the positive correlation between the gas 
metallicity and the outflow is that the gas with a high metallicity (and with a high dust 
abundance consequently) has a larger optical depth. Therefore such metal-rich clouds 
are more easily affected by the outward radiative pressure. 
Also, the resonant scattering could contribute to the outflow.
If the resonant scattering exerts radiation pressure, higher metallicity gas has higher optical depth in the resonance
scattering, that leads to more powerful outflow.
Interestingly, our results for low-redshift ($z < 0.4$) QSOs and those for high-redshift ($1.7 < z < 4$) QSOs 
show a similar trend, even though these two studies examine completely different 
cosmic epochs. This suggests that the physics of the gas accretion onto the SMBH 
and its relation to the nuclear star formation is universal through the cosmic timescale.
 
On the other hand, the SMBH mass does not show a significant correlation with the
outflow. The correlation coefficients are 0.25 for low-redshift QSOs and -0.39 for 
high-redshift QSOs \citep{Wang2011}. This implies that the SMBH mass itself is not 
an important factor for nuclear activities such as nuclear star formation and outflow.
This interpretation is consistent to the recent work by \citep{Shin2013}, reporting that 
there is no statistically meaningful correlation between the SMBH mass and the BLR 
metallicity.\\

\subsection{OIII shift}\label{OIII}
The powerful AGN outflow is sometimes seen also in the NLR. \cite{Komossa2008} 
showed that the \OIII\ emission (a relatively high-ionization NLR line) in some AGNs
shows a blue shift with respect to other low ionization narrow lines, such as \SII. 
Since it is known that high-ionization NLR clouds (emitting \OIII) locate more inner
parts in the NLR than low-ionization NLR clouds emitting such as \SII, \OI, and \OII\ 
\citep{Ferguson1997,Nagao2001a}, it is naturally expected that high-ionization NLR
lines more frequently show outflow features than low-ionization NLR lines \citep[e.g.,][]{Bae2014}. Then a 
question naturally arises --- is there any physical connection between BLR outflows 
and NLR outflows? In terms of the gas metallicity, some earlier works reported the 
physical connection between gas clouds in NLRs and BLRs \citep[e.g.,][]{Fu2007,
Du2014}. In these works, it is inferred that the outflowing BLR gas affects the chemical
property in the NLR. Therefore, it is highly interesting to examine possible physical link
between the BLR outflow and NLR outflow. This can be studied by investigating the 
outflow indices for \CIV\ and those for \OIII.

To investigate a possible link between the BLR outflow and NLR outflow, we focus 
on the outflow indices for \OIII\ and compare them with the outflow indices for \CIV\ 
in our low-redshift AGN sample.
Figure 5 shows the results of the comparison of VSI between \OIII\ and \CIV. Here
we derive the VSI by adopting different references for the systemic redshift; \Hb, \OI, 
\OII, and \SII, for checking the robustness of the inferred results. 
There is no apparent correlation between the VSI of \OIII\ and \CIV, regardless of 
the adopted redshift reference. The Spearman's rank-order correlation test also 
suggests that there is not statistically significant correlation between them (the exact
results of the rank-order correlation test are given in Figure 5).
However the obtained results do not necessarily mean that the BLR outflow and NLR
outflow is independent, because the sample size of our low-redshift AGNs is so small
that the covered range of the \OIII\ VSI is very limited (from --150 km s$^{-1}$ to
+250 km s$^{-1}$ roughly). 
The studies of \OIII\ in the literature typically focus on the wing component, which represents the non-gravitational 
kinematic signature, finding evidence of strong outflows \citep{Woo2016,Bae2016,Woo2017}. On the other hand the velocity shift
of \OIII\ is reported to be relatively small \citep{Bae2014, Woo2016}. Thus, our comparison of BLR and NLR outflows based on the velocity shift
may not reveal a strong trend. 
The comparison of the outflow kinematics between \OIII\ and 
\CIV\ for much larger samples with detailed line profile analysis will be an interesting test to understand a possible link
between the BLR and NLR outflows.

\begin{figure}{}
\includegraphics[width = 0.48\textwidth]{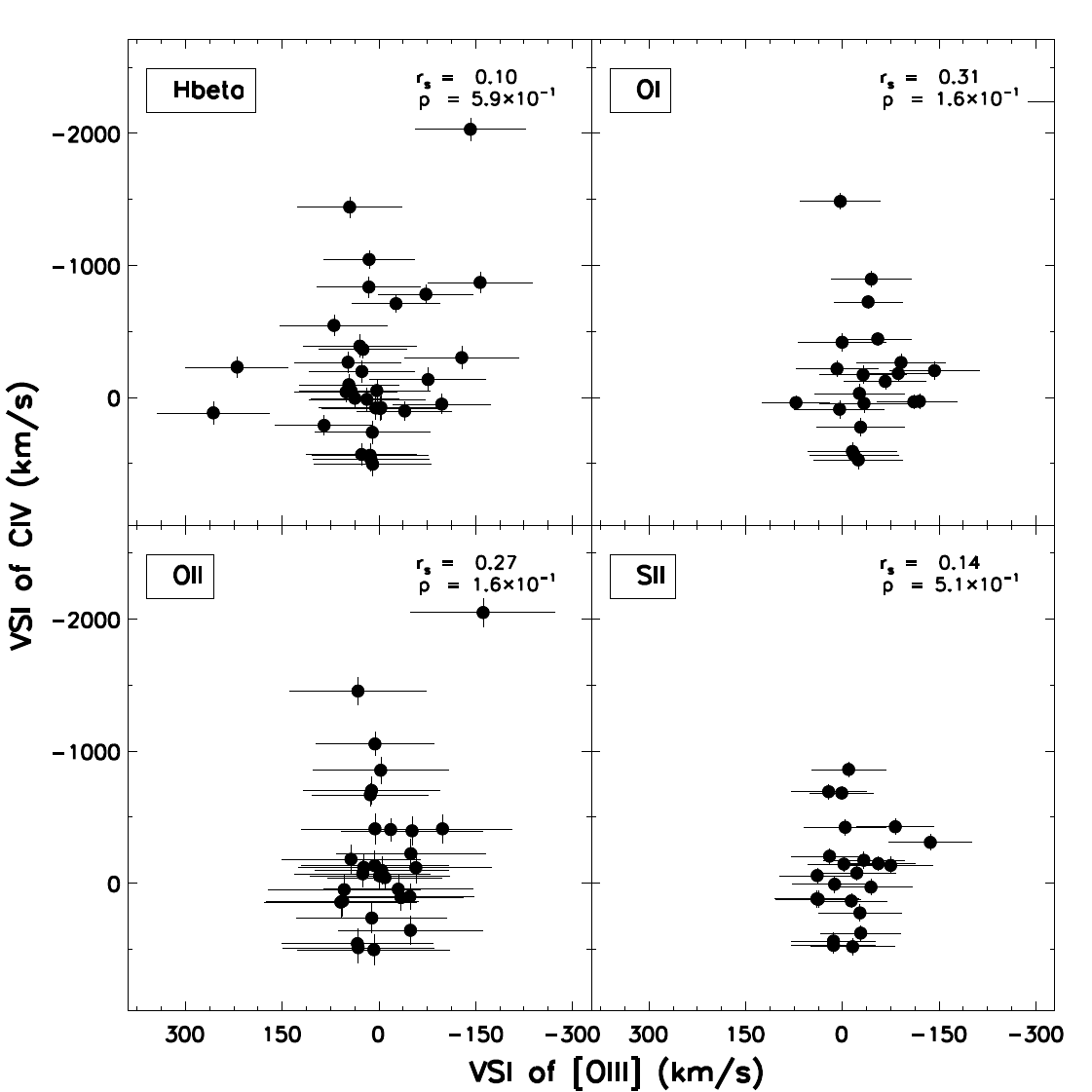}
\caption{
     Comparison between \OIII\ and \CIV\ VSIs. Each panel shows the
     comparison of them, adopting different narrow line for determining the systemic 
     redshift (top left: \Hb, top right: \OI, bottom left: \OII, and bottom right: \SII).
\label{fig:allspec1}
}
\end{figure}

\section{Summary}\label{summary}
To understand the AGN gas outflow at the BLR of low-redshift AGNs, we derive 
AGN outflow indicators of 34 low-redshift AGNs by analyzing the \CIV\ velocity profile
and by adopting the systemic redshift defined by some low-ionization narrow emission 
lines such as \SII\ and \OI. By comparing the measured outflow indicators with 
various AGN properties, the following results are obtained.

1. The outflow indicators (VSI and BAI of \CIV) show weak positive correlations with the
Eddington ratio, bolometric luminosity, and BLR metallicity indicators, suggesting
that there is a connection among the past star formation, accretion, and AGN outflow
in the nuclear region of host galaxies. However, there are little correlation between the BLR
outflow indicators and the SMBH mass.

2. The inferred relation among the BLR metallicity, Eddington ratio, and BLR outflow,
seen in our low-redshift AGN sample, is consistent with that seen in high-redshift
QSO sample \citep{Wang2011,Wang2012}. This implies there is no significantly 
cosmological evolution of the mechanism triggering the AGN activity.

3. A possible relation between the BLR outflow and NLR outflow is also investigated,
by comparing the outflow indicator (VSI) for \CIV\ and \OIII. However, any apparent
correlation between the two is not identified. This may be due to the small size of our
sample, suggesting that more extensive studies based on larger samples are
required.\\

\acknowledgements
We would like to thank the anonymous referee for helpful comments which improved the clarity of the paper. 
This work was supported 
%by the Korea Astronomy and Space Science Institute (KASI) grant funded by the Korea government (MEST). J.-H.W. acknowledges the support 
by the National Research Foundation of Korea (NRF) grant funded by 
the Korea government (MEST) (No. 2016R1A2B3011457 and 2010-0027910). T.N. acknowledges the 
support by JSPS (grant no. 25707010, 16H01101, and 16H03958) and 
by the JGC-S Scholarship Foundation. The Mikulski Archive for Space Telescopes 
(MAST) is a NASA funded project to support and provide a variety of astronomical data archives to the astronomical 
community, with the primary focus on scientifically related data sets in the optical, 
ultraviolet, and near-infrared parts of the spectrum. MAST is located at the Space 
Telescope Science Institute (STScI). 
This research has partly made use of the NASA/IPAC Extragalactic Database 
(NED) which is operated by the Jet Propulsion Laboratory, California Institute of 
Technology, under contract with NASA. Funding for the SDSS 
has been provided by the Alfred P. Sloan Foundation, the Participating Institutions, 
the National Science Foundation, the U.S. Department of Energy, the National 
Aeronautics and Space Administration, the Japanese Monbukagakusho, the Max 
Planck Society, and the Higher Education Funding Council for England. 
The SDSS Web Site is http://www.sdss.org/.

%\bibliography{outflow}

\end{document}